\documentclass[12pt,dvipdfmx]{article}

\usepackage{a4}
\usepackage{a4wide}

\usepackage{amsmath}
\usepackage{amscd}
\usepackage{amsfonts}
\usepackage{amssymb}
\usepackage{mathrsfs}
\usepackage{fancybox} 
\usepackage{enumerate}

\usepackage{cite}
\usepackage{bm}
\usepackage{amscd}
\usepackage{graphicx}
\usepackage[dvips]{color}
\usepackage{epsfig}

\makeatletter

\@addtoreset{equation}{section}
\makeatother

\usepackage{amsthm} 

\usepackage{ulem}

\def\det{\mathrm{det}}
\def\det{\mathrm{det}}

\def\half{{1\over2}}

\def\={\stackrel{\bullet}{=}}

\def\({\left(}
\def\){\right)}
\def\[{\left[}
\def\]{\right]}

\def \be {\begin{equation}}
\def \ee {\end{equation}}
\def \beqa {\begin{eqnarray}}
\def \eeqa {\end{eqnarray}}
\def \beal#1 {\begin{align}#1\end{align}}
\def \bes#1 {\begin{equation}\begin{split}#1\end{split}\end{equation}}
\def \nn {\notag\\}

\makeatletter

\@addtoreset{equation}{section}
\makeatother

\begin{document}

\begin{titlepage}
\title{
\vspace{-2cm}
\begin{flushright}
\normalsize{ 
YITP-20-24 \\ 
OU-HET 1044
}
\end{flushright}
       \vspace{1.5cm}
What does a quantum black hole look like?
       \vspace{1.cm}
}
\author{
 Sinya Aoki\thanks{saoki[at]yukawa.kyoto-u.ac.jp },
\; Tetsuya Onogi\thanks{onogi[at]phy.sci.osaka-u.ac.jp},\; 
Shuichi Yokoyama\thanks{shuichi.yokoyama[at]yukawa.kyoto-u.ac.jp},\; 
\\[25pt] 
${}^{*}{}^{\dagger}{}^{\ddagger}$ {\normalsize\it Center for Gravitational Physics,} \\
{\normalsize\it Yukawa Institute for Theoretical Physics, Kyoto University,}\\
{\normalsize\it Kitashirakawa Oiwake-cho, Sakyo-Ku, Kyoto 606-8502, Japan}
\\[10pt]
${}^\dagger$ {\normalsize\it Department of Physics, Osaka University,}\\ 
{\normalsize\it Toyonaka, Osaka 560-0043, Japan,}
}

\date{}

\maketitle

\thispagestyle{empty}


\begin{abstract}
\vspace{0.3cm}
\normalsize

We consider a free theory of multiple scalar fields at finite temperature and study
the induced  geometry defined through a free flow of the scalar fields,
following the method proposed by the present authors as a possible candidate of the constructive approach for AdS/CFT correspondence. We find that the holographic metric has the following properties: i) It is an asymptotic Anti-de Sitter (AdS) black brane metric with some unknown matter contribution. ii) It has no coordinate singularity and milder curvature singularity. iii) Its time component decays exponentially at a certain AdS radial slice. 
We find that the matter spreads all over the space, which we speculate to be due to thermal excitation of infinitely many massless higher spin fields. We conjecture that the above three are generic features of a black hole holographically realized by the flow equation method.

\end{abstract}
\end{titlepage}

\section{Introduction} 
\label{introduction} 

A black hole is a key object to build a bridge between general relativity and quantum field theory. 
It was seminally shown that a black hole entails a horizon and a singularity \cite{Penrose:1964wq,Hawking:1966vg,Hawking:1969sw} and behaves as a thermodynamical object \cite{Hawking:1971tu,Bekenstein:1973ur,Bardeen:1973gs} accompanied with particle radiation in a Planck distribution \cite{Hawking:1974sw,Hartle:1976tp}. 
Hawking argued that the existence of a singularity causes breakdown of a fundamental law of physics \cite{Hawking:1976ra}. For instance, suppose a black hole formed from heavy matter. The black hole radiates particles carrying only the thermodynamic information, while it gradually evaporates losing more detailed information of its initial state, which is a non-unitary transition from a pure state to a mixed one.
This information loss puzzle has been investigated actively up to the present developing various new ideas and techniques (See \cite{Page:1979tc,Banks:1983by,tHooft:1990fkf,Russo:1992yh,Verlinde:1993sg,Susskind:1993if,Stephens:1993an,Polchinski:1994zs} for earlier studies.)  

An innovative method to investigate general relativity with 
quantum effects taken into account 
is the AdS/CFT correspondence \cite{Maldacena:1997re,Witten:1998qj,Gubser:1998bc}. In AdS, a black hole stably exists \cite{Hawking:1982dh}, and the information loss puzzle can be analyzed from a dual conformal field theory (CFT)  \cite{Lowe:1999pk,Jacobson:1999mi,Maldacena:2001kr}.
One of the keys to solve the puzzle is the resolution of the black hole singularity by the quantum effects of gravity. It was argued that the singularity indeed can be resolved by summing over geometries around the saddle points of the path integral, which also restores the unitarity \cite{Maldacena:2001kr,Polchinski:1994zs}. The resolution of the black hole singularity as well as some  coordinate singularity such as the  horizon in a quantum gravity may be natural from the viewpoint of string theory, in which black holes consist of branes and the microscopic degrees of freedom carried by the black hole are accounted for by strings ending on the branes
\cite{Strominger:1996sh,Callan:1996dv,Strominger:1997eq}. In this realization a black hole may be a fuzzy object with no apparent horizon \cite{Lunin:2001jy} (See also \cite{David:2002wn,Mathur:2005zp}.)

A novel approach to realize the framework of holography has been proposed and developed by the authors of the present letter, in which a 'holographic' direction is conjectured to emerge by a flow equation 
\cite{Aoki:2015dla,Aoki:2016ohw,Aoki:2016env,Aoki:2017bru}. 
A virtue of this approach is that it is applicable to a wide class of quantum field theories incorporating traditional techniques of quantum field theories such as the $1/N$ expansion. 
The flow equation approach also enables us to study classical and quantum aspects of gravity on several classic backgrounds \cite{Aoki:2017uce,Aoki:2018dmc,Aoki:2019bfb,Aoki:2019dim,Yokoyama:2020tgs}. 

The purpose of this paper is to apply the flow equation method to a finite temperature system and study its induced geometry, which is supposed to be described by a black hole or a black brane solution if the flow method correctly describes the holography. (See also \cite{Gursoy:2018umf}.) As a first step we study a free theory with multiple scalar fields at finite temperature. The dual gravity theory is conjectured to be a free higher spin theory consisting of all even spin fields \cite{Klebanov:2002ja}, which is known to admit a black hole solution in four dimensions \cite{Didenko:2009td,Iazeolla:2011cb,Iazeolla:2017vng}. 

The rest of this letter is organized as follows. 
In Sec.~\ref{metric}, we consider multiple free scalar fields and smear them by the free flow equation. Using the 2-point function of the flowed field  at finite temperature we compute  the bulk holographic metric. We study its asymptotic behaviors at the UV and the deep IR and make comparison with those for the AdS black hole.
In Sec.~\ref{EMT}, we calculate the matter energy momentum tensor (EMT) from the bulk metric through the Einstein equation, and discuss its properties, 
in particular, the behavior of the EMT near the boundary.
Our conclusion and discussion are given in  Sec.~\ref{Discussion}.

\section{Holographic geometry at finite temperature
} 
\label{metric} 

\subsection{Propagators and holographic metric} 

We begin with a multiple free scalar theory on ${\mathbb R}^d$, and flow the scalar fields by a free flow: 
\be 
{\partial \over \partial t}\phi^a(x;t) = \partial^2 \phi^a(x;t), ~~~ \phi^a(x;0)=\varphi^{a}(x),
\ee
where $\partial^2=\partial_\mu\partial_\mu$ with $\mu=1,\cdots,d$, $\varphi^{a}(x)$ is a original massless scalar field with $a=1,2,\cdots, N$.
As derived in Ref.~\cite{Aoki:2017bru}, the 2-point function of the flowed field  becomes
\beqa
\langle \phi^a(x;t) \phi^b(y;s) \rangle_0 
= \frac{\delta^{ab}}{[4(t+s)]^{{d-2\over2}}\Gamma({d-2\over2})} F_0\left(\frac{(x-y)^2}{t+s}\right),
\label{eq:2pt_phi}
 \eeqa
where $F_0(u) =\displaystyle \int_0^1 dv\, v^{d/2-2} e^{-\frac{u}{4} v}$.
We assume $d>2$ to avoid the divergence of the integral, which corresponds to the bad infrared behavior of a massless scalar at $d=2$.

In order to study the system at temperature $T$, we compactify one of the directions denoted by $x^0$, so that we set the periodic boundary condition for each scalar field in the $x^0$ direction with the periodicity $1/T$.
Then the 2-point function at finite temperature can be obtained by summing over the 'images' produced by the compactification 
\beqa
\langle \phi^{a}(x^0,\vec x;t) \phi^{b}(y^0,\vec y;s) \rangle_T 
&=&\sum_{n=-\infty}^{\infty} \langle \phi^{a}(x^0,\vec x;t) \phi^{b}(y^0+n/T,\vec y;s) \rangle_0 .
\eeqa
Since this 2-point function of the flowed field has no contact singularity, we can normalize the smeared field using the 2-point function at zero temperature as
\beqa 
\sigma^{a}(x^0,\vec x;t) &=&\frac{\phi^{a}(x^0,\vec x;t)}{\sqrt{\langle \phi^2(x^0,\vec x;t)\rangle_{0}}},  
\eeqa
where $\phi^2 =\sum_{a=1}^N\phi^a \phi^a$. 
Employing this normalized field we define a holographic metric by
\beqa
g_{MN} (X) &=& \ell^2
\sum_{a=1}^N
\langle \partial_M \sigma^{a}(x^0,\vec x;t) \partial_N \sigma^{a}(x^0,\vec x;t) \rangle_T,
\label{eq:metric}
\eeqa  
which can be interpreted as an information metric\cite{Aoki:2017bru},
where $\ell$ is a length scale fixed by hand, and $(X^M)=(x^0,\vec x,\tau)$ with $\tau =\sqrt{2d t}$.
Although our proposal that eq.~(\ref{eq:metric})  can be a constructive definition to realize the metric in the holographic geometry is still a conjecture, it indeed successfully reproduces the bulk AdS metric at $T=0$\cite{Aoki:2017bru}.
For the present case at finite temperature, the metric in eq.~(\ref{eq:metric})
 can be calculated as follows.
\beal{
g_{00}(X) =& \frac{ L_{\rm AdS}^2}{\tau^2} \frac{d}{2} \[ F(d,z) - z\frac{d}{dz} F(d,z)\]
, \\
g_{\tau\tau}(X) =& \frac{L_{\rm AdS}^2}{\tau^2 }\[\frac{d-2}{2} F(d-2,z)
-\half z\frac{d}{dz} F(d-2,z)+\frac14\(z\frac{d}{dz}\)^2 F(d-2,z) \], \\
g_{ij}(X) =& \delta_{ij} \frac{L_{\rm AdS}^2}{\tau^2} \frac{d}{2} F(d,z), 
}
where $z=\tau T$ is a dimensionless quantity corresponding to the AdS radial coordinate, $L_{\rm AdS}^2 =\ell^2 (d-2)/2,$
\beal{
F(s, w) =& \int_0^1dv\, v^{s/2-1}\, \theta_3\(e^{-\frac{dv}{4z^2}}\),
\label{F}
}
with the elliptic theta function  $\theta_3(q):= 1 + 2\sum_{n=1}^\infty q^{\half n^2} $.

\subsection{Asymptotic behaviors}

Let us study the asymptotic behaviors of the holographic metric. 
To this end we introduce useful expressions of the function $F$ defined by \eqref{F}.
An expression good for small $z$ region is
\beqa
F(s,z) &=& \frac{2}{s} + 2\({4z^2\over d}\)^{\frac{s}{2}} \Gamma\left(\frac{s}{2}\right)\zeta(s) - \delta F_{\rm UV}(s,z),
 \eeqa
where
\beqa
\delta F_{\rm UV}(s,z) = 2\({4z^2\over d}\)^{s/2} \sum_{n=1}^\infty n^{-s}\Gamma\left(\frac{s}{2},\frac{d}{4z^2}n^2\right) 
~~~~~~~~~~~~~~
\label{eq:FA1}
\eeqa
with the incomplete Gamma function $\Gamma(s,a)$, which is exponentially small for large $a$.
For large $z$, on the other hand, using the Poisson summation formula,
$\theta_3(e^{-x}) = \sqrt{\dfrac{\pi}{x}}\theta_3\(e^{-\pi^2/x}\right )$,
the expression becomes
\beqa
F(s,z) &=& \sqrt{{\pi\over d}}\frac{4z}{s-1}+ \delta F_{\rm IR}(s,z), 
\eeqa
where
\beqa
\delta F_{\rm IR}(s,z) = 4 \sqrt{{\pi\over d}}z \sum_{n=1}^\infty
\left(4\pi^2 n^2 {z^2\over d}\right)^{\frac{s-1}{2}} \Gamma\left( \frac{1-s}{2}, 4\pi^2n^2 {z^2\over d} \right).~~~~~~
\eeqa
Note that $\delta F_{\rm UV/IR} (d,z)$ damp exponentially for small/large $z$.

We write the metric in a standard form such that 
\beqa
g_{00} (X) =  \frac{L_{\rm AdS}^2}{\tau^2}f_0(z), \quad
g_{\tau\tau}(X) = \frac{L_{\rm AdS}^2}{\tau^2}f_\tau(z),\quad
g_{ij} (X) = \delta_{ij} \frac{L_{\rm AdS}^2}{\tau^2}f_i(z). ~~~~
\eeqa
In the small $z$ region, we have
\beqa
f_0(z) &=& 1 - z^d  (d-1)A_d  -  \frac{d}{2} \(1 - z\frac\partial{\partial z} \) \delta F_{\rm UV} \(d, z\),
\label{eq:f1_UV} \\
f_\tau(z)&=& 1 + z^{d-2}\frac{(d-2)}{2} A_{d-2} -\(
\frac{(d-2)}{2} - \frac{1}{2}z \frac\partial{\partial z}  +\frac14 \( z\frac\partial{\partial z} \)^2  \)
\delta F_{\rm UV}(d-2, z),~~~~~~
\label{eq:f2_UV}\\
f_i(z) &=& 1+  z^d  A_d -\frac{d}{2}\delta F_{\rm UV}\(d, z\),
\label{eq:f3_UV}
\eeqa
where $A_s :=(4/d)^{\frac{s}{2}} s\, \Gamma(s/2) \zeta(s)$.
Note that the metric in the small $z$ limit  describes the AdS space. 

In the large $z$ region, we obtain
\beqa
f_0(z) &=&\frac{d}{2}\(1 -  z\frac\partial{\partial z} \)\delta F_{\rm IR}(d,z) 
 \label{eq:f1_IR}\\
f_\tau(z) &=& z \sqrt{\frac{4\pi}{d}} \(\frac{2d-5}{2d-6}\) + \half\left(
(d-2)
- z\frac\partial{\partial z} +\half\(  z\frac\partial{\partial z}\)^2  \right) \delta F_{\rm IR}(d-2,z), 
 \label{eq:f2_IR}\\
f_i(z) &=& z \sqrt{\frac{4\pi}{d}}\(\frac{d}{d-1}\) + \frac{d}{2}\delta F_{\rm IR}(d,z) .
 \label{eq:f3_IR}
\eeqa
Remark that we need $d>3$ for large $z$ to avoid the divergence in $F_{\rm IR}(d-2,z)$, which corresponds to an infrared singularity of a 3-dimensional massless scalar field at finite $T$, whose modes with zero Matsubara frequency  behave like 2-dimensional massless scalar.
Thus the metric in the large $z$ limit becomes
\beqa
g_{00} (X) = 0, \quad
g_{\tau\tau}(X) = L_{\rm AdS}^2\sqrt{\frac{4\pi}{d}}\(\frac{2d-5}{2d-6}\)\frac{T}{\tau},
\quad
g_{ij} (X) = \delta_{ij}L_{\rm AdS}^2\sqrt{\frac{4\pi}{d}}\(\frac{d}{d-1}\)\frac{T}{\tau}.\notag
\eeqa

From the dependence on $z$, we find that at finite temperature the metric asymptotically remains AdS with the same cosmological constant at small $z$,
while it is deformed towards larger $z$.
What is this new spacetime?

\subsection{Comparison with AdS Blackhole}
\begin{figure}[htb]
\begin{center}
\includegraphics[width=0.45\hsize, angle=0]{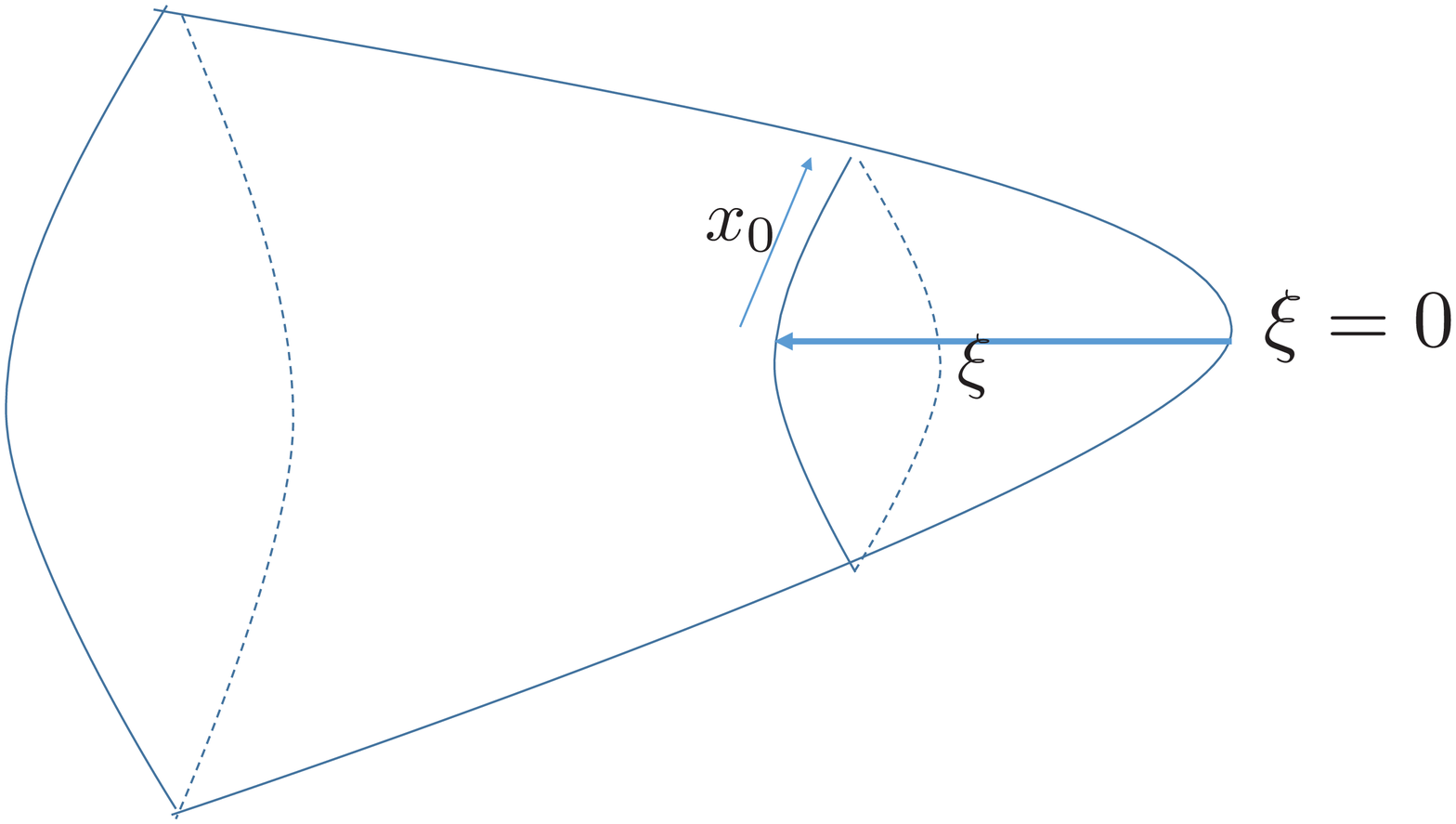}
\hspace{1cm}
\includegraphics[width=0.4\hsize, angle=0]{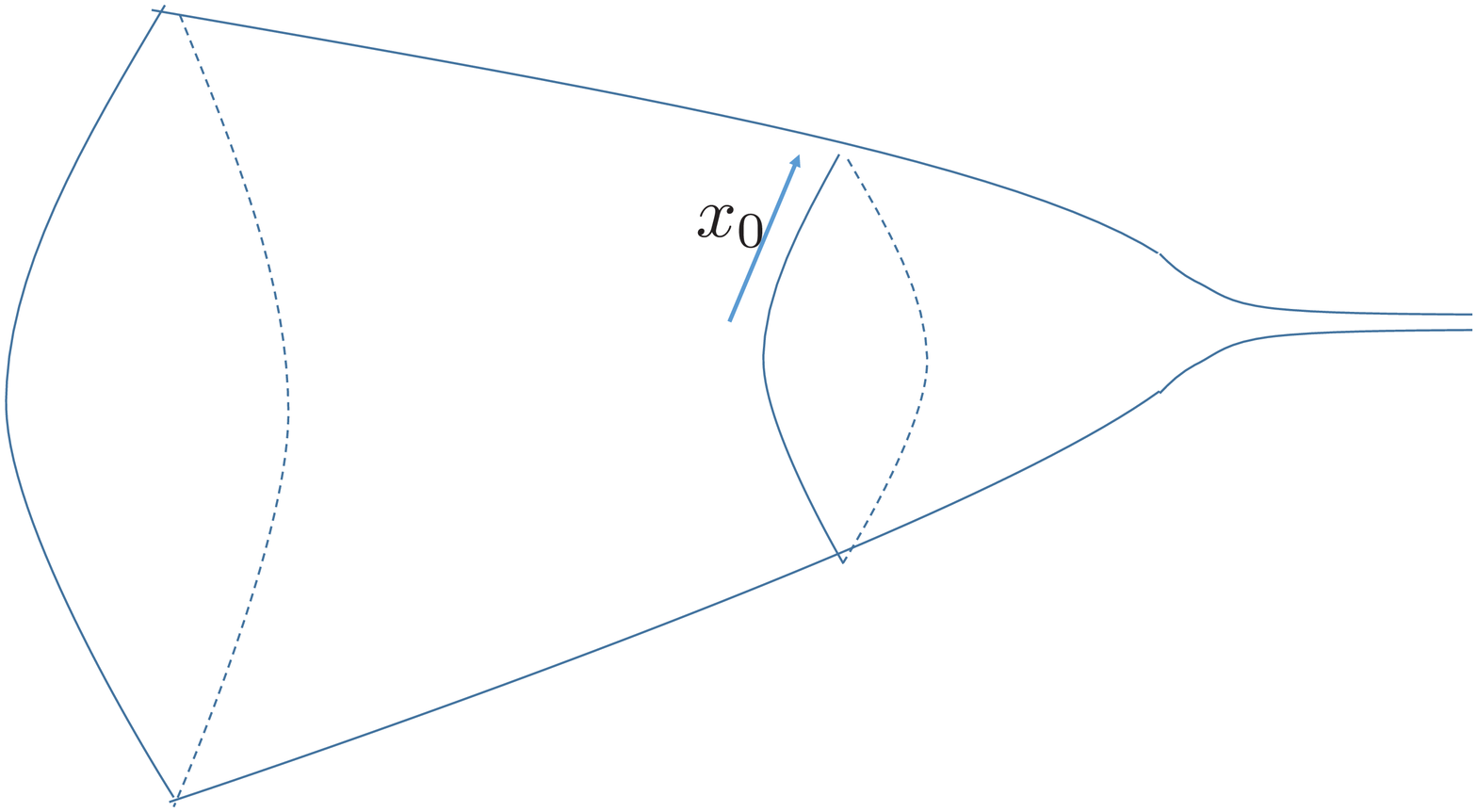}
\end{center}
\caption{Left panel: The total space of the Euclidean AdS blackhole forms a cigar-like structure near the horizon. 
The tip of the cigar is $\xi=0$ and $\xi$ is the 'distance" from the tip while $x^0$ is the 'angle'.
Right panel:  Large part of the total space of our system forms a cigar-like structure except for the region near the tip, which becomes a thin long tube stretching to $z=\infty$. 
}
\label{fig:Cigar}
\end{figure}

For comparison, let us write down the metric of the 
Euclidean AdS Blackhole.
\begin{eqnarray}
ds^2 = \frac{L_{\rm AdS}^2}{\tau^2}
\left(f_\tau^{\rm BH}(\tau) d\tau^2
+f_0^{\rm BH}(\tau) (dx^0)^2
+\sum_{i=1}^{d-1} f_i^{\rm BH}(\tau) (dx^i)^2
 \right),
 \label{AdSBH}
\end{eqnarray}
where 
\begin{eqnarray}
f_\tau^{\rm BH}(\tau) = \left(1-\frac{\tau^d}{\tau_0^d}\right)^{-1}, 
&
f_0^{\rm BH}(\tau) = 1-\displaystyle{\frac{\tau^d}{\tau_0^d}}, 
&
f_i^{\rm BH}(\tau) =1   \  (i=1,\cdots, d-1)
\end{eqnarray}
with $\tau_0$ being the inverse horizon radius. 
Near the horizon, $\tau = \tau_0 ( 1-\xi^2)$ with $\xi\ll 1$, we have
\begin{eqnarray}
ds^2 \approx \frac{L_{\rm AdS}^2}{\tau_0^2} \left(\frac{4\tau_0^2}{d}  (d\xi)^2 + d \cdot \xi^2(dx^0)^2 + 
\sum_{i=1}^{d-1} (dx^i)^2\right).
\end{eqnarray}
The  absence of the conical singularity determines the temperature as $T =\frac{d}{2\tau_0}$.
This expression tells us that the total space forms a cigar-like structure near the horizon
at which the space ends as shown in the left panel of Fig.\ref{fig:Cigar}.
The tip of the cigar is $\xi=0$ and $\xi$ is the 'distance" from the tip while $x^0$ is the 'angle'. at the horizon. 
Thus, in Euclidean AdS blackhole, one finds that one can only cover the region outside the horizon.

Now let us come back to our system. Our total space shares some of the features of 
the AdS Blackhole. First, the metric is asymptotically AdS and at becomes exactly AdS at $T=0$. Also, our  system asymptotically approaches $g_{00} =0$, which  may be interpreted   as a ``horizon" of the blackhole (or more precisely blackbrane in this case). However, there are also differences. Since $g_{00}$ never goes to zero at finite $\tau$, the whole space can reach $\tau=\infty$. 
To see these differences quantitatively, 
we numerically evaluate 
eqs.~(\ref{eq:f1_UV}), (\ref{eq:f2_UV}), (\ref{eq:f3_UV})  for small $z$ or their dual expressions  eqs.~(\ref{eq:f1_IR}), (\ref{eq:f2_IR}), (\ref{eq:f3_IR}) for large $z$,
by replacing  the infinite sum of incomplete  Gamma functions with 
the finite sum, which gives negligible errors as long as one uses the proper expression out of the two.
\begin{figure}[htb]
\begin{center}
\includegraphics[width=0.8\hsize]{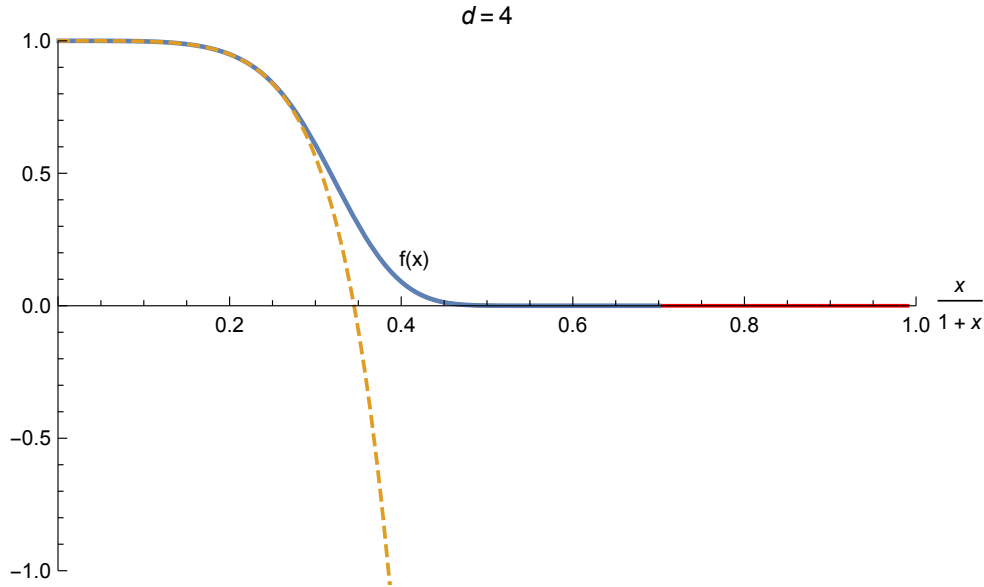}
\end{center}
\caption{A solid line is $f_0(z)$ as a function of $z/(1+z)$ at $d=4$.
The dashed line is the NLO behavior at small $\tau T$, which agrees with $g_{00}$ for the AdS blackbrane.
}
\label{fig:Metric4d}
\end{figure}
Fig.~\ref{fig:Metric4d} shows $f_0(z)$ as a function of $z/(1+z)$ at $d=4$, where
0(1) in the $x$-axes corresponds to $\tau=0 (\infty)$.
In the figure, a blue solid line from 0 to 0.7 in the $x$-axes
is evaluated using eq.~(\ref{eq:f1_UV}),
while a red one from 0.4 to  1.0 ( a part between 0.4 and 0.7 is masked by the blue line) 
is obtained by  eq.~(\ref{eq:f1_IR}),  with a sum of $n$ up to 20 for each case.
Both agree well between 0.4 and 0.7, showing that truncation errors for the infinite summation is well under control. An orange dashed line is the next-to-leading (NLO) approximation of $f_0$ at small $z$, given by
\beqa
f_0(z) \simeq 1 - (d-1)  A_d \, z^d, 
\eeqa
which has the same functional form in  $z$ of the AdS blackhole (or blackbrane). 
Our bulk metric $g_{00}$ deviates from the one for the AdS blackhole around 0.3, 
becomes almost zero around 0.45 and exponentially small beyond 0.45. Note that $z=1$
corresponds to 0.5 of the horizontal axis in the figure. 
Although the real ``horizon" does not appear in this metric, the effective (or pseudo) horizon seems to exist around $z/(1+z) \simeq 0.45$ ($z \simeq 9/11$).

We thus find that  the $g_{00}$ component of the bulk metric in the small $z$ region has a qualitatively similar behavior as that of AdS blackbrane.
Large part of the total space looks like a cigar-like structure.
However, the region near the tip i.e. $0.45 \leq \tau z/(1+\tau z)\leq 1$ becomes a thin long tube stretching to $z=\infty$ 
where  the $0$-th direction shrunk into  a circle with an exponentially small radius as shown in the right panel of Fig.\ref{fig:Cigar}. Therefore,  
it would be hard to distinguish whether this thin long tube region exists or not from an external observer.

By the way, the fact that $g_{00}(z) \simeq 0$ at sufficiently large $z$ can be naturally understood from the boundary field theory point of view as follows. The flow smears the boundary field with the smearing length  $\tau$. Therefore, if $\tau \simeq 1/T$, the smearing reaches the temporal boundary, so that
no more information in the temporal direction implies $g_{00}\simeq 0$.
In other words, the dimensional reduction due to the temperature $T > 1/\tau$ produces the blackbrane-like object in the bulk geometry.  

Our bulk metric shows deviations from the blackbrane-like object even for small $\tau$ regime. Indeed  the NLO
approximation of $f_\tau$ and $f_i$ at small $\tau$ become
\beqa
f_\tau(z) &\simeq& 1+ \frac{(d-2)}{2} A_{d-2}\, z^{d-2}, \quad
f_i(z) \simeq 1+ A_d z^d.
\label{eq:f_tau}
\eeqa
By the change of variables for $\tau$ as $\tilde{\tau}=\tau(1-\frac{1}{2}A_d\tau^d)$, we have
\beqa
\tilde f_0 \simeq 1 - d A_d \tilde z^d, \quad
\tilde f_{\tilde\tau} \simeq {1 \over \tilde f_0} + \frac{(d-2)}{2} A_{d-2} \, \tilde z^{d-2},
\quad
\tilde f_i \simeq  1,
\label{eq:BH-like}
\eeqa
where $\tilde z=\tilde{\tau} T$. 
Without the second term of $f_\tau$, this describes nothing but the AdS blackbrane \eqref{AdSBH}.
The second term of $f_\tau$ has stronger effect than others near the boundary. 
Since this term overwhelms the contribution from the AdS blackbrane, one finds that the metric cannot be the solution of the vacuum Einstein equation with cosmological term even far outside (small $z$).

Where does this new effect come from? One possible interpretation could be the matter effect. 
This is not so surprising, since the thermal excitations of the massless scalar field may 
give significant contributions to 
produce this effect, due to the absence of energy gap between the vacuum and excited states. 
Correspondingly, gases of massless excited states may appear in the bulk.
Another interpretation can be the deviation from Einstein gravity.  Indeed if we take the free $O(N)$ vector model at the boundary, the bulk theory is expected to correspond to the free higher spin theory with all even spin.
In future, we would like to address an interpretation of $\tau^{d-2}$ effect in the metric more explicitly. 
Note that if we interpret the higher spin fields as exotic matter fields, both the first and the second scenarios can be regarded as 'Einstein gravity with new matter effect'. Therefore in the next section, we extract the energy momentum tensor from the new matter and study its property assuming the Einstein equation.

\subsection{Comments on entropy}
Before closing this section, we comment on entropy of the dual holographic space, which is expected to match the entropy of the boundary theory \cite{Strominger:1996sh,Callan:1996dv,Strominger:1997eq}.  

For computation of entropy in the bulk, denoted by $S_{\rm bulk}$,
we use the metric \eqref{eq:BH-like}, in which the effect of matter hanging over the space is 
forced to appear only in  the radial component by a coordinate transformation. 
By choosing an AdS radial slice at which the UV approximation is valid, the metric \eqref{eq:BH-like} asymptotically behaves as an AdS blackhole, whose horizon is located at $\tilde\tau=\tilde\tau_H$ with $\tilde\tau_H T = 1/(dA_d)^{1/d}$. 
We compute the bulk entropy $S_{\rm bulk}$ by employing  the Beckenstein-Hawking entropy formula. 
\beal{
S_{\rm bulk} = {A \over 4 G_{d+1}},
}
where $G_{d+1}$ is the $d+1$ dimensional Newton constant and $A$ is the area of the horizon 
at $\tilde\tau=\tilde\tau_H$  given by 
\begin{eqnarray}
A &=& \int_{\tilde\tau=\tilde\tau_H} d^{d-1} x\sqrt{\det g_{ij} }
= V \left(\frac{L_{\rm AdS}^2 \tilde f_i(\tilde\tau_H T)}{\tilde\tau_H^2}\right)^{\frac{d-1}{2}}, 
\end{eqnarray}
where $V\equiv \int d^{d-1}x$. 
Note that the second term in $\tilde f_{\tilde \tau}$, which is absent for the AdS blackhole,
does not affect the result when the Beckenstein-Hawking formula is applied.

On the other hand, the entropy of the massless scalar field on the boundary is computed as 
\begin{eqnarray}
S_{\rm bdry} &=& 2 N \pi^{-\frac{d}{2}} \Gamma\left(\frac{d}{2}+1\right)\zeta(d) V T^{d-1}, \quad
\end{eqnarray}
Therefore, the ratio of the two entropies becomes
\begin{eqnarray}
\frac{S_{\rm bdry}} {S_{\rm bulk}}
=C { N G_{d+1} \over L_{\rm AdS}^{d-1}}, ~~~ 
\label{entropyRatio}
\end{eqnarray}
where $C=\frac{ (\pi^{-1} d)^{d\over2} }{d\, 2^{d-2} \tilde z_H \tilde f_i(\tilde z_H) ^{\frac{d-1}{2}}} $ with $\tilde z_H=(dA_d)^{-1/d}$.
If we approximate $\tilde f_i(\tilde z_H)\simeq 1$, then $C\simeq 0.207$ at $d=4$. 
The ratio ${ N G_{d+1} / L_{\rm AdS}^{d-1}}$, which is of the order $1$ according to the AdS/CFT correspondence, can be independently determined  from other information such as the comparison of correlation functions.  
A consistency check whether the entropy ratio \eqref{entropyRatio} becomes unity is left for future work.

\section{Energy momentum tensor}
\label{EMT}
In this section, we consider the matter energy momentum tensor (EMT), defined from the metric 
through the Einstein tensor as $
\kappa^2 T_{AB}:=G_{AB} + \Lambda g_{AB}$, 
where $\Lambda = -d(d-1)/(2 L_{\rm AdS}^2)$ and
$\kappa^2$ is the Newton constant. 
In terms of $f_{0,\tau,i}$, we obtain
\beqa
\kappa^2 T^0{}_{0} &=&\frac{1}{L_{\rm ADS}^2} \frac{1}{f_\tau}\left[
 \frac{d(d-1)}{2}(1-f_\tau)
 -\frac{d-1}{2}  \left\{-\log f_\tau +(d-1)\log f_i\right\}_{\tau} +\frac{d-1}{2}\left\{\log f_i\right\}_{\tau\tau}  \right. \nn 
&+&
\left. \frac{d-1}{4} \{\log f_i \}_\tau \left\{  - \log f_\tau 
+\frac{d}{2}\log  f_i \right\}_\tau
\right], \\
\kappa^2 T^\tau{}_{\tau} &=&\frac{1}{L_{\rm ADS}^2}\frac{1}{f_\tau}\[\frac{d(d-1)}{2}(1-f_\tau) -\frac{d-1}{2}  \left\{\log f_0  + (d-1)\log f_i\right\}_{\tau}\right. \nn
&+&\left. \frac{d-1}{4} \{\log f_i \}_\tau \left\{   \log f_0
+\frac{d-2}{2} \log  f_i  \right\}_\tau\], 
\eeqa
\beqa
\kappa^2 T^i{}_{j} &=&\frac{1}{L_{\rm ADS}^2}\delta^i_{j} \frac{1}{f_\tau}\left[
 \frac{d(d-1)}{2}(1-f_\tau)
 -\frac{d-1}{2}  \left\{ \log\left(\frac{f_0}{f_\tau}\right) +(d-2)\log f_i\right\}_{\tau}\right. \nn
&  +&\frac{1}{2}\left\{\log f_0 +(d-2) \log f_i\right\}_{\tau\tau}  
+  
\frac{d-2}{4} \{\log f_i \}_\tau \left\{   \log \(\frac{f_0}{f_\tau}\)
+\frac{d-1}{2} \log  f_i  \right\}_\tau\nn
&+&\left.\ \frac{1}{4}\left\{\log f_0\right\}_\tau\left\{\log\left(\frac{f_0}{f_\tau}\right)\right\}_\tau
\right] ,
\eeqa
where $\{X\}_\tau:= \tau \partial_\tau X$ and $\{X\}_{\tau\tau}:=\tau^2 \partial_\tau^2 X$.

At small $z$, we have
\beqa
\kappa^2 T^0{}_0(z) &\simeq & -(d-1) B z^{d-2}, \qquad B :=\frac{d-2}{2} \frac{A_{d-2}}{L_{\rm AdS}^2}, \\
\kappa^2 T^\tau{}_\tau(z) &\simeq& -\frac{d(d-1)}{2} B z^{d-2}, \quad
\kappa^2 T^i{}_j (z) \simeq  -\delta^i_j (d-1)  B z^{d-2} ,
\eeqa
which are dominated by $f_\tau$ in eq.~(\ref{eq:f_tau}) and their derivatives.
At large $z$, we obtain
\beqa
\kappa^2 T^0{}_0(z) &\simeq& -\frac{1}{L_{\rm AdS}^2}\frac{d(d-1)}{2}\(1 -  \frac{(d-3)(d+2)}{4 (2d-5) \sqrt{\pi d}}\frac{1}{z}
 \),
\\
\kappa^2 T^\tau{}_\tau(z) &\simeq&\frac{1}{L_{\rm AdS}^2} \frac{2(d-1)(d-3)\pi^2}{(2d-5) \sqrt{\pi d}} z,
\quad
\kappa^2 T^i{}_j (z) \simeq \delta^i_j \frac{1}{L_{\rm AdS}^2}\frac{16(d-3)\pi^4}{d(2d-5)\sqrt{\pi d}} z^3 . 
\label{eq:TzzTii}
\eeqa
\begin{figure}[htb]
\begin{center}
\includegraphics[width=0.8\hsize]{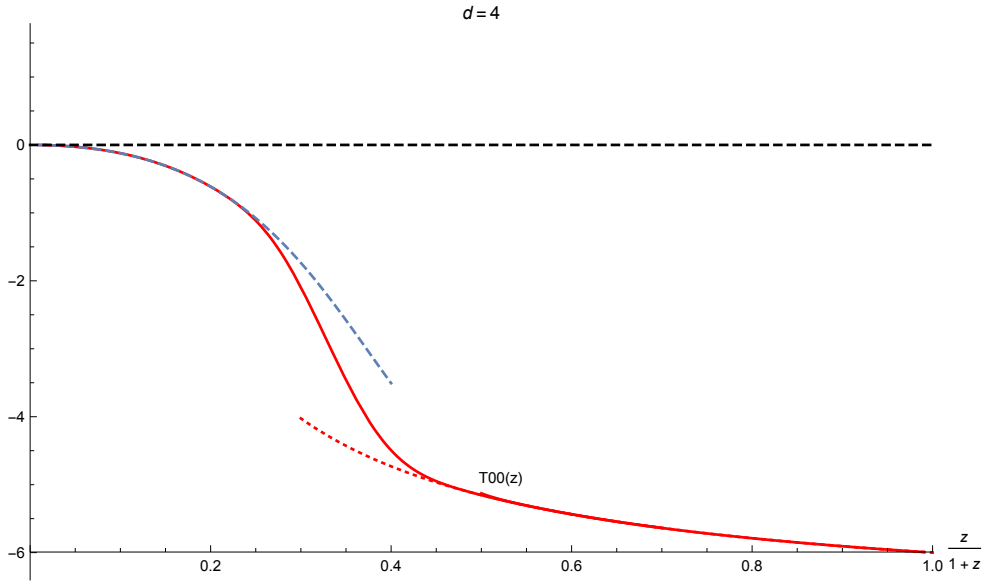}
\end{center}
\caption{The red solid line represents $ L_{\rm AdS}^2 \kappa^2 T^0{}_{0}(z)$ as a function of $z/(1+z)$ at $d=4$. The blue dashed (red dotted) line is the NLO behavior at small (large) $z$. 
}
\label{fig:T004d}
\end{figure}
Fig.~\ref{fig:T004d} shows $L_{\rm AdS}^2 \kappa^2 T^0{}_{0} (z)$ (blue and red solid lines) as a function of $z/(1+z)$  at $d=4$.
As seen in the figure,
$L_{\rm AdS}^2 \kappa^2 T^0{}_{0} (z)$ is non-zero  everywhere but its absolute value reaches the maximum,  $d(d-1)/2$, at $z=\infty$. 
Comparing it with the energy momentum tensor of the AdS blackbrane solution, which vanishes except the singularity and diverges at that point, we observe that the singularity formed by coalescence of matter is resolved and matter spreads all over the space. 
\begin{figure}[htb]
\begin{center}
\includegraphics[width=0.8\hsize]{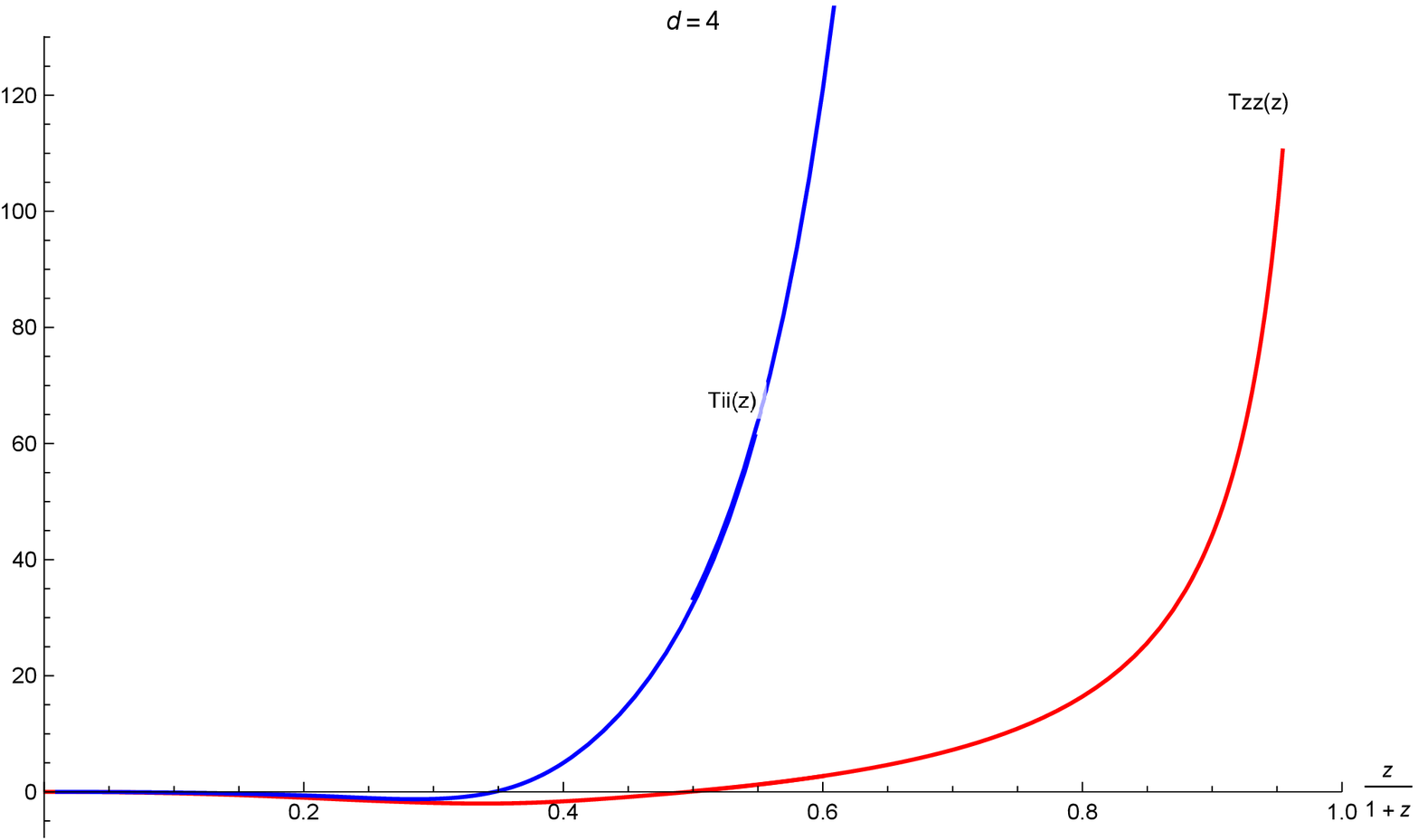}
\end{center}
\caption{The red (blue) solid line represents $L_{\rm AdS}^2 \kappa^2 T^\tau{}_{\tau}(z)$ 
($ L_{\rm AdS}^2 \kappa^2 T^i{}_{i}(z)$) as a function of $z/(1+z)$ at $d=4$. 
}
\label{fig:TzzTii4d}
\end{figure}
On the other hand,
Fig.~\ref{fig:TzzTii4d} shows $L_{\rm AdS}^2 \kappa^2 T^\tau{}_{\tau}(z)$ (red solid line) and
$ L_{\rm AdS}^2 \kappa^2 T^i{}_{i}(z)$ (blue solid line) as a function of $z/(1+z)$.
As shown in eq.~(\ref{eq:TzzTii}), 
$T^\tau{}_{\tau}(z)$ and $ T^i{}_{i}(z)$ diverge as $z$ and $z^3$, respectively, in the large $z$ limit. However, compared to that of the AdS blackhole, this divergence is also suppressed, and the corresponding singularity does not lead to divergence for a global physical quantity, 
as is defined by its integration over the space with a weight of a determinant factor $\sqrt g = {\sqrt{f_z f_0 f_i^{d-1}}L^{d+1}/\tau^{d+1}}$ \cite{Aoki:2020prb}.  

While the behavior of the dual geometry becomes milder in the IR, it gets more singular in the UV ( or small $z$ ) region. We presume that this will be due to the effect of matter spread over the space, which would clump at the singularity in the case of the AdS blackhole.
Since the $O(N)$ free massless scalar theory at the boundary is expected to be dual to the higher spin theory in the bulk, thermal effects at the boundary can easily excite massless higher spin fields in the bulk. This may cause 
the non-standard behaviors of EMTs as $z^{d-2}$ near $z = 0$.
In fact, even taking the spin zero contribution alone, namely massless free scalar field with conformal coupling in the bulk, 
it gives $z^{d-1}$ contributions to EMTs at $z\simeq 0$.
Although there is a mismatch of the power of $z$ only for this contribution, 
it is suggestive that the extra infinitely many massless fields in the bulk gives the non-standard contribution to the EMT.  
In future studies, it would be interesting to see whether infinitely many massless higher spin fields 
generate such $z^{d-2}$ behaviors of EMTs near the boundary.

\section{Discussion} 
\label{Discussion} 

We have investigated a conjectured holographic geometry of a free O($N$) vector theory at finite temperature by the flow equation approach. 
The resulting metric behaves as an asymptotic AdS black brane with some matter hanging all over the space though it is free from the coordinate singularity as well as that of the matter energy momentum tensor $T^0{}_0$. We observed that 
other components of the energy momentum tensor have milder singularity at the IR, which does not lead to divergence for a global quantity. Assuming the known higher spin/vector model duality we presume that the unknown matter contribution to the energy momentum tensor comes from infinitely many massless higher spin fields excited by thermal effects. 

The holographic metric obtained in this paper has remarkable features stated above and in the abstract as well. We strongly suspect that these features remain unchanged even if interactions are tuned on, as long as they are weak enough. In other words, these features will change only when the system becomes strongly coupled enough. More precisely, 
as a coupling constant in the CFT side becomes stronger, the matter
spreading over the entire space in the free case gradually
turns to clump around the deeper IR region, and in the strongly
coupled limit the matter collapses to form a horizon. 
In this limit, we expect the Cosmic censorship hypothesis to be fully recovered \cite{Penrose:1964wq,Hawking:1966vg,Hawking:1969sw}, though it presumably works in the asymptotic region without taking the limit. 
In this sense it is highly important to extend this work to the current system including interactions and Yang-Mills theories and test whether the above picture is correct or not.  

In this letter we exclude a two dimensional case for a general analysis of free theories. It would be interesting to extend this analysis to a two dimensional interacting CFT at finite temperature, which has also been proposed to have a dual higher spin theory \cite{Gaberdiel:2010pz}.
It is known that three dimensional higher spin theories admit not only a black hole with conventional global charges \cite{Didenko:2006zd} but also one with higher spin charges \cite{Kraus:2011ds,Gutperle:2011kf}.
(See also \cite{Kraus:2006wn,Perez:2012cf,Ammon:2012wc,David:2012iu,deBoer:2013gz}.)  
These black holes in three dimensions are peculiar in the respect that for the former there is no curvature singularity \cite{Banados:1992wn,Banados:1992gq} and for the latter the horizon becomes gauge-dependent \cite{Ammon:2011nk} (See also \cite{Castro:2011fm,Bunster:2014mua}.) 
Holography provides a tool to study these black holes from a dual CFT viewpoint \cite{Gaberdiel:2012yb}. 

We hope to come back to these issues in the near future. 

\section*{Acknowledgment}

We would like to thank Drs. Shigeki Sugimoto and Tadashi Takayanagi for useful discussions and valuable comments.
This work is supported in part by the Grant-in-Aid of the Japanese Ministry of Education, Sciences and Technology, Sports and Culture (MEXT) for Scientific Research (Nos.~JP16H03978,  JP18K03620, JP18H05236, JP19K03847). 
S. A. is also supported in part  
by a priority issue (Elucidation of the fundamental laws and evolution of the universe) to be tackled by using Post ``K" Computer, and by Joint Institute for Computational Fundamental Science (JICFuS).
T. O. would like to thank YITP for their kind hospitality during his
stay for the sabbatical leave from his home institute.

\bibliographystyle{utphys}
\bibliography{FiniteT}

\end{document}